\documentclass[a4paper,10pt,twoside]{cpc-hepnp}

\usepackage{multicol}
\usepackage{graphicx}
\usepackage{booktabs}
\usepackage{amssymb,bm,mathrsfs,bbm,amscd}
\usepackage[tbtags]{amsmath}
\usepackage{lastpage}

\newcommand{\pdHeh}{\mbox{$pd\to{}^3\mathrm{He}\,\eta$}}
\newcommand{\E}[1]{\mbox{$\times$10$^{#1}$}}

\begin{document}

\fancyhead[co]{\footnotesize M. B\"uscher: Meson-production experiments at COSY}

\footnotetext[0]{Received \today}

\title{Meson-production experiments at COSY-J\"ulich}

\author{%
      M. B\"uscher$^{1;1)}$\email{m.buescher@fz-juelich.de}%
}
\maketitle

\address{%
  1~(Institute f\"ur Kernphysik and J\"ulich Center for Hadron
  Physics, Forschungszentrum J\"ulich, 52425 J\"ulich Germany) }

\begin{abstract}
  Selected results from experiments at COSY-J\"ulich are presented: an
  attempt to measure the mass of the $\eta$ meson with high precision
  (ANKE facility), first steps towards the detection of rare $\eta$
  decays (WASA), and several measurements of $K\bar K$-pair
  production (ANKE, COSY-11, MOMO).
\end{abstract}

\begin{keyword}
Meson production
\end{keyword}

\begin{pacs}
14.40.Cs Other mesons with $S=C=0$, mass $< 2.5\,$GeV
\end{pacs}

\begin{multicols}{2}

\section{Introduction}

The Cooler Synchrotron and storage ring COSY\cite{maier1997} of the
Forschungszentrum J\"{u}lich delivers unpolarized and polarized beams
of protons and deuterons with momenta up to 3.7 GeV/c.  Currently,
these are utilized at three large experimental facilities, ANKE and
WASA at internal target stations and COSY-TOF at the extracted beam.

\begin{itemize}
\item {\bf ANKE} (Apparatus for Studies of Nucleon and Kaon Ejectiles)
  is a large acceptance forward magnetic
  spectrometer\cite{barsov2001}. The central dipole is movable to
  adjust the momenta of the detected particles independent of the beam
  momentum. Using hydrogen or deuterium cluster
  targets\cite{khoukaz1999}, close-to-threshold reactions on protons
  or neutrons can be measured.  In addition, a polarized internal
  target with a storage cell can be used. The ANKE detectors have been
  optimized for charged kaon detection\cite{Buescher:2002zc} which
  allow one to identify $K^+K^-$ pairs\cite{Hartmann:2007ks} in
  coincidence with fast forward going protons and
  deuterons\cite{anke-fwd}.

\item {\bf TOF} (Time Of Flight) is a non-magnetic spectrometer
  combining excellent tracking capabilities with large acceptance and
  full azimuthal symmetry allowing to measure complete Dalitz
  plots\cite{ElSamad:2009yu}.  TOF is optimized for final states with
  strangeness, see {\em e.g.\/} Ref.\cite{AbdelSamad:2006qu}. With the
  new straw tube tracking system (STT), TOF will have a significantly
  improved mass resolution and reconstruction efficiency.

\item {\bf WASA} (Wide Angle Shower Apparatus), is an almost 4$\pi$
  spectrometer for neutral and charged particles. It comprises an
  electro-magnetic calorimeter, a very thin superconducting solenoid,
  inner and forward trigger and tracking detectors, and a
  frozen-pellet target.  WASA was build at TSL Uppsala and it was used
  there until 2005 at the CELSIUS ring\cite{Bargholtz:2008ze}. The
  system was transferred to COSY in 2005 and it was operational
  already after one year. One of the main goals for WASA at COSY is
  the study of rare $\eta$-meson decays involving photons and
  electrons\cite{Adam:2004ch}.
\end{itemize}

Here we report on selected ongoing measurements carried out at the ANKE
and WASA facilities and on data from the decommissioned COSY-11
and MOMO experiments.  For other results from COSY see the
contributions to this conference by M.~Bashkanov (two-pion production)
and H.~Machner ($\eta$ mesons in matter).

\section{Precision measurement of the $\eta$ mass}
Measurements of the $\eta$-meson mass at different experimental
facilities over the last decade have resulted in remarkably precise
results. However, these differ by up to 500~keV/$c^2$, which is more
than 8 standard deviations of the quoted individual experimental
uncertainties. As a consequence the PDG tables no longer consider
experiments based on the identification of the $\eta$ as a
missing-mass peak of a hadronic reaction. In order to check whether
there is an intrinsic problem in such experiments, a refined
measurement of the $dp\to\mathrm{^3He}\,\eta$ reaction has been
carried out at COSY.

After producing the $\eta$ mesons through the
$dp\to\mathrm{^3He}\,\eta$ reaction, the $^3$He is detected with the
ANKE forward detectors\cite{anke-fwd}. Reactions with the production
of an $\eta$ meson can then be unambiguously identified from a
missing-mass criterion\cite{mersmann2007}.  Due to the simple two-body
kinematics of the $\mathrm{^3He}\,\eta$ final state, the $\eta$ mass
can be determined through a thorough reconstruction of the size of the
$^3$He kinematic ellipse. This, in turn, relies on a precise
identification of the reaction threshold and an accurate measurement
of the associated beam momentum\cite{Goslawski:2009vf}.

The measurements were carried out at thirteen beam energies in the
range $1<Q<10$~MeV as well as $Q = -5$~MeV for background
studies\cite{Goslawski:2009vf}. To determine the $\eta$ mass with a
precision that is competitive to other state-of-the-art measurements,
{\em i.e.\/} $\Delta m_{\eta}<50$~keV/$c^2$\cite{pdg2008}, the
associated beam momenta have to be fixed with an accuracy of $\Delta
p /p < 10^{-4}$.  This can be achieved via the precise measurement of
the frequency of depolarizing resonances $f_r$ and of the beam
revolution frequency $f_0$. The usefulness of the technique relies on
the fact that frequencies can be routinely measured to $10^{-5}$.

From the measured frequencies $f_r$ and $f_0$ the kinematical factor
$\gamma=E/mc^2$, and hence the beam momentum $p$, can be deduced:
\begin{eqnarray}
  \nonumber
  \gamma &=& \frac{1}{G_d} \left( \frac{f_r}{f_0} - 1 \right) \\
  \nonumber
  p &=& m_d \, \beta \, \gamma = m_d \; \sqrt{\gamma^2 - 1}\ \ .
\end{eqnarray}
Here $E$ denotes the deuteron total energy, $m_d$ the deuteron mass
and $G_d=(g-2)/2$ is the gyromagnetic anomaly of the deuteron, where
$g$ is the gyromagnetic factor.

The spins of the polarized deuteron beam particles precess around the
normal to the plane of the machine, which is generally horizontal.
The spin can be perturbed by the application of a horizontal
\emph{rf} magnetic field from, for example, a solenoid.  The beam
depolarizes when the frequency of the externally applied field
coincides with that of the spin precession in the ring.  An example
of a spin-resonance spectrum is shown in
Fig.~\ref{fig:spin_resonance}. Far away from the spin resonance at
1.0116~MHz and 1.0120~MHz, a high degree of polarization was measured
using the EDDA detector\cite{EDDA} as a beam polarimeter. In
contrast, when the solenoid frequency coincides with that of the spin
precession $f_r$, the beam is maximally depolarized.

\begin{center}
  \includegraphics[width=0.95\linewidth]{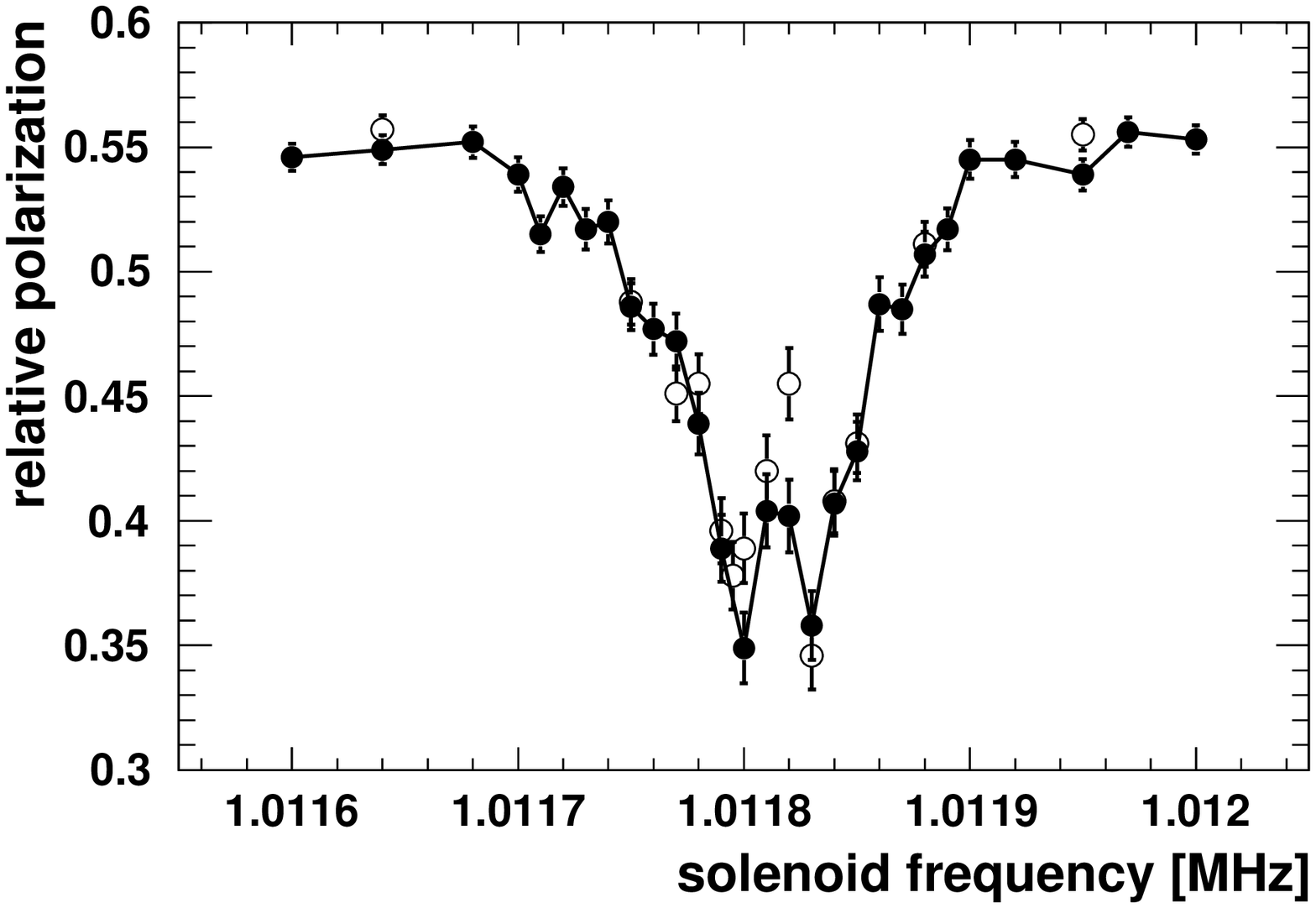}
  \figcaption{\label{fig:spin_resonance}Spin resonance measurements at
    one beam momentum (closed circles). The open symbols represent results
    obtained for an extended cycle time, where the perturbing solenoid
    was switched on after 178~s.}
\end{center}

The beam revolution frequency $f_0$ was measured from the Schottky
noise of the deuteron beam. The origin of this effect is the
statistical distribution of the charged particles in the beam. This
leads to random current fluctuations, which induce a voltage signal at
a beam pick-up. The Fourier transform of this voltage-to-time signal
by a spectrum analyzer delivers the frequency distribution around the
harmonics of the revolution frequency of the beam.

The accuracies to which both of the frequencies are determined are
dominated by systematic effects. The revolution frequency measured by
the Schottky spectrum analyzer has an uncertainty of $\Delta f_0 =
6$~Hz, corresponding to one in the beam momentum of $50$~keV/$c$. The
error in the determination of the spin resonance arises from the small
variations of the orbit length and $\Delta f_r = 15$~Hz corresponds to
an uncertainty in the beam momentum of $164$~keV/$c$. Because these
systematic uncertainties are independent, they are added quadratically
to give a total uncertainty $\Delta p/p \leqslant 6\times 10^{-5}$,
{\em i.e.\/}, a precision of $170$~keV/$c$ for beam momenta in the
range of $3100-3200$~MeV/$c$.  This is over an order of magnitude
better than ever reached before for a standard experiment in the COSY
ring.  This accuracy will ultimately allow the mass of the $\eta$
meson to be measured with a precision of $\Delta m_{\eta} \leqslant
50$~keV/$c^2$.  The remaining part of the data analysis is under way.

\section{Towards rare $\eta$ decays}
The initial stage of the $\eta$ meson decay experiments with WASA was
carried out at CELSIUS using the $pp\to pp\eta$ and
$pd\to{}^3\mathrm{He}\,\eta$ reactions close to threshold. Results on
the Dalitz plot density for the $\eta \to 3\pi^0$ decay
\cite{Bashkanov:2007iy} and on the branching ratios of some leptonic
decay channels \cite{Bargholtz:2006gz,Berlowski:2008zz} have been
reported.

During the last two years, larger data samples have been collected at
COSY on $\eta$ decays from $pp$ and $pd$ collisions. From a short
production run in April 2007 120k events have been extracted for the
$\eta \to 3\pi^0$ Dalitz plot\cite{Adolph:2008vn}.  In the next 4- and
8-week run periods in fall 2008 and fall 2009, $\eta$ decays were
studied using the $pd\to{}^3\mathrm{He}\,\eta$ reaction at a beam
energy of 1~GeV.  Unbiased data samples of $1.1\times 10^7$ and $\sim
2\times 10^7$ $\eta$ meson decays were collected, respectively.  In
addition, a few shorter runs were also carried out with the aim of
optimizing the conditions for $\eta$ meson production from $pp$
collisions.

\begin{center}
  \includegraphics[width=0.49\linewidth]{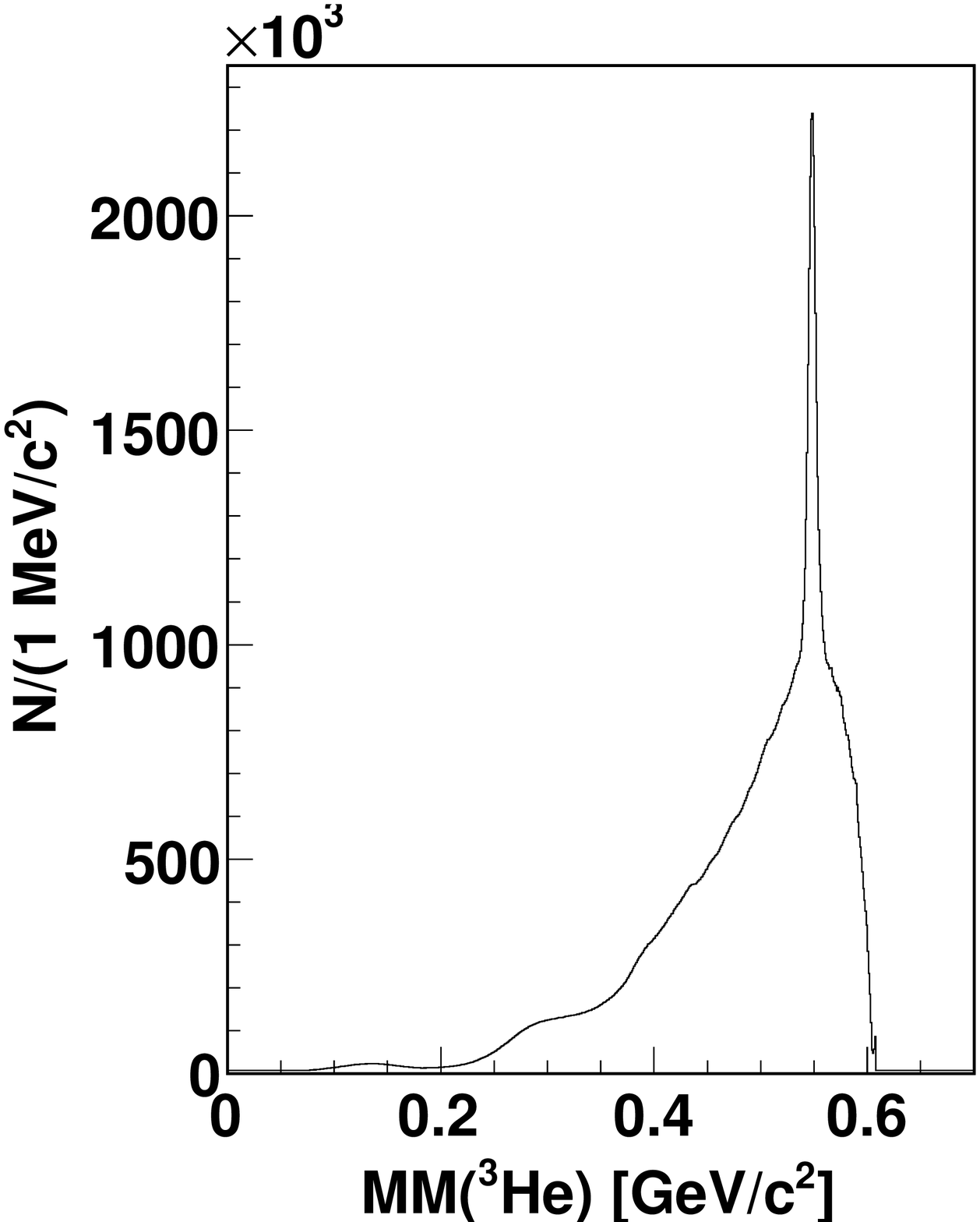}
  \includegraphics[width=0.49\linewidth]{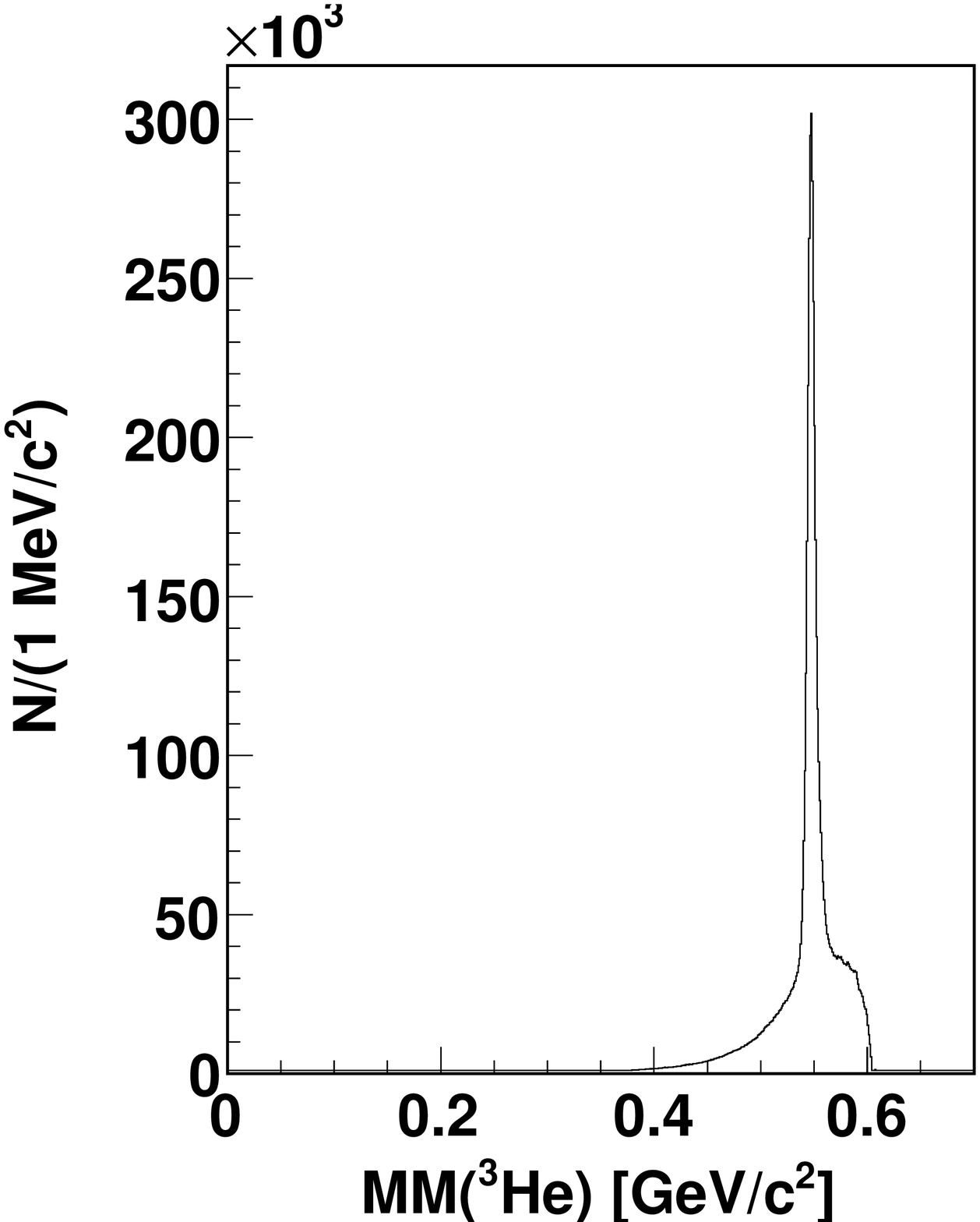}
  \figcaption{\label{fig:mm3he} Distribution of $^3$He missing mass
    for the $pd\to^3$He$X$ reaction at 1.0~GeV. The trigger was based
    entirely on $^3$He signal in the forward detector without any bias on the decay system.
    The plot shows all the data collected during the 2008 run period.
    (Left) data analysis based only on the forward detector track --
    there are about 1.1\E{7} events in the peak at the $\eta$ meson
    mass. (Right) in addition two-photon invariant masses $\ge$300
    MeV/c$^2$ have also been demanded.}
\end{center}

The cross section of \pdHeh\ reaction rises quickly and reaches a
plateau value of 400~nb already at 2~MeV above the threshold.  The
requirement of a $^3$He ion in the final state selects only a very
tiny fraction of the total $pd$ cross section (about 0.1\%). The
separation of $^3$He from protons and deuterons is quite reliable and
is implemented on the trigger level.  When doing this, one obtains
trigger rates at maximum luminosity that are well within the DAQ
capabilities without imposing any additional constraints on the
$\eta$ decay pattern.

The \pdHeh\ reaction is tagged by identifying the $^3$He particles in
the WASA forward detector, which covers scattering angles from
3$^\circ$ to 18$^\circ$.  Figure \ref{fig:mm3he} shows the missing
mass of the reconstructed $^3$He at 1.0 GeV beam energy. In the left
panel all data collected in the 2008 $pd$ run period are shown.  In
the right panel after selection of a particular decay channel.

The WASA detector was optimized for the measurement of
electron-positron pairs and photons from the decays of $\pi^0$ and
$\eta$ mesons.  The performance of the detector can be checked by
studying single Dalitz decays $\eta\to e^+ e^- \gamma $ and $\pi^0\to
e^+ e^- \gamma$.  The decays were observed in both $pp$ and $pd$
interactions.  Figure~\ref{fig:eeg} shows an example of the $\eta\to
e^+ e^- \gamma $ decay identification in the invariant mass of $e^+
e^- \gamma$ from the $pd$ data.  The data from both $pd$ and $pp$
runs should contain a few thousands of $\eta \to e^+ e^- \gamma$
events.

\begin{center}
  \includegraphics[width=\linewidth]{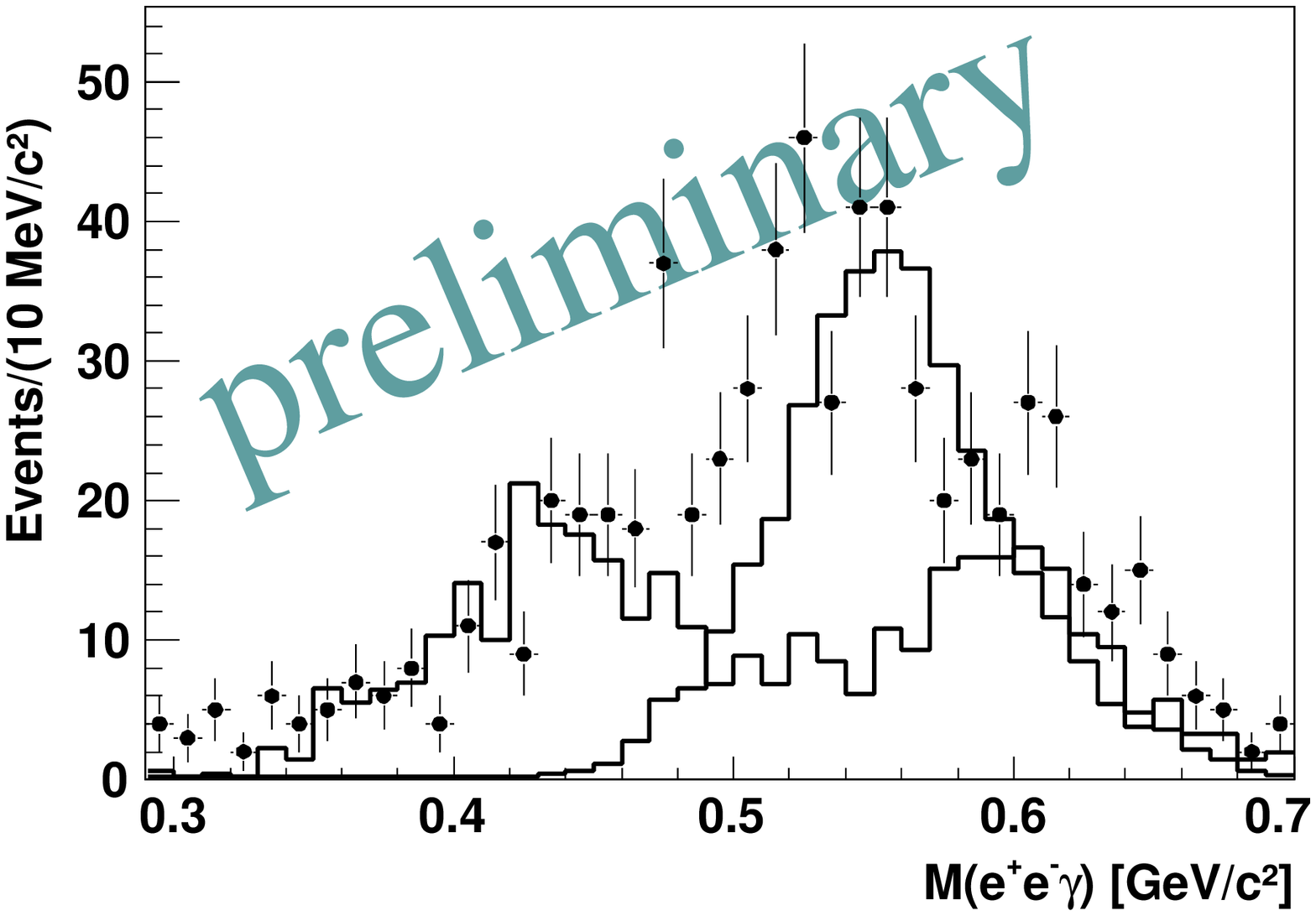}
  \figcaption{Single Dalitz decay $\eta\to e^+ e^-\gamma$ events
    for about 10\% of the 2008 $pd$ data. The histograms represent
    Monte Carlo simulations of the signal (solid line) and background
    (dashed line).}
  \label{fig:eeg}
\end{center}

The decay $\eta\to \pi^+\pi^-e^+e^-$ allows precise tests of the
chiral anomaly through the comparison with existing ChPT
calculations. A high statistics measurement of $\eta\to
\pi^+\pi^-e^+e^-$ will also provide constraints for a new kind of
flavor-conserving CP violation by measuring the asymmetry of the
dihedral angle between the pion- and electron planes. The present PDG
value for the branching ratio of the $\eta \to \pi^+\pi^-e^+e^-$
decay is 4.2\E{-4}.

The status of the analysis of the $\eta \to \pi^+\pi^-e^+e^-$ decay
is shown in Fig.~\ref{fig:pipiee}.  A clear signature with about 300
events of the $\eta\to \pi^+\pi^-e^+ e^-$ decay is seen in the $^3$He
missing mass. The line represents a polynomial fit to the direct pion
production reactions. According to the Monte Carlo simulations, 50\%
of the events in the $\eta$ mass peak are from $\eta\to\pi^+\pi^-e^+
e^-$.

\begin{center}
  \includegraphics[width=\linewidth]{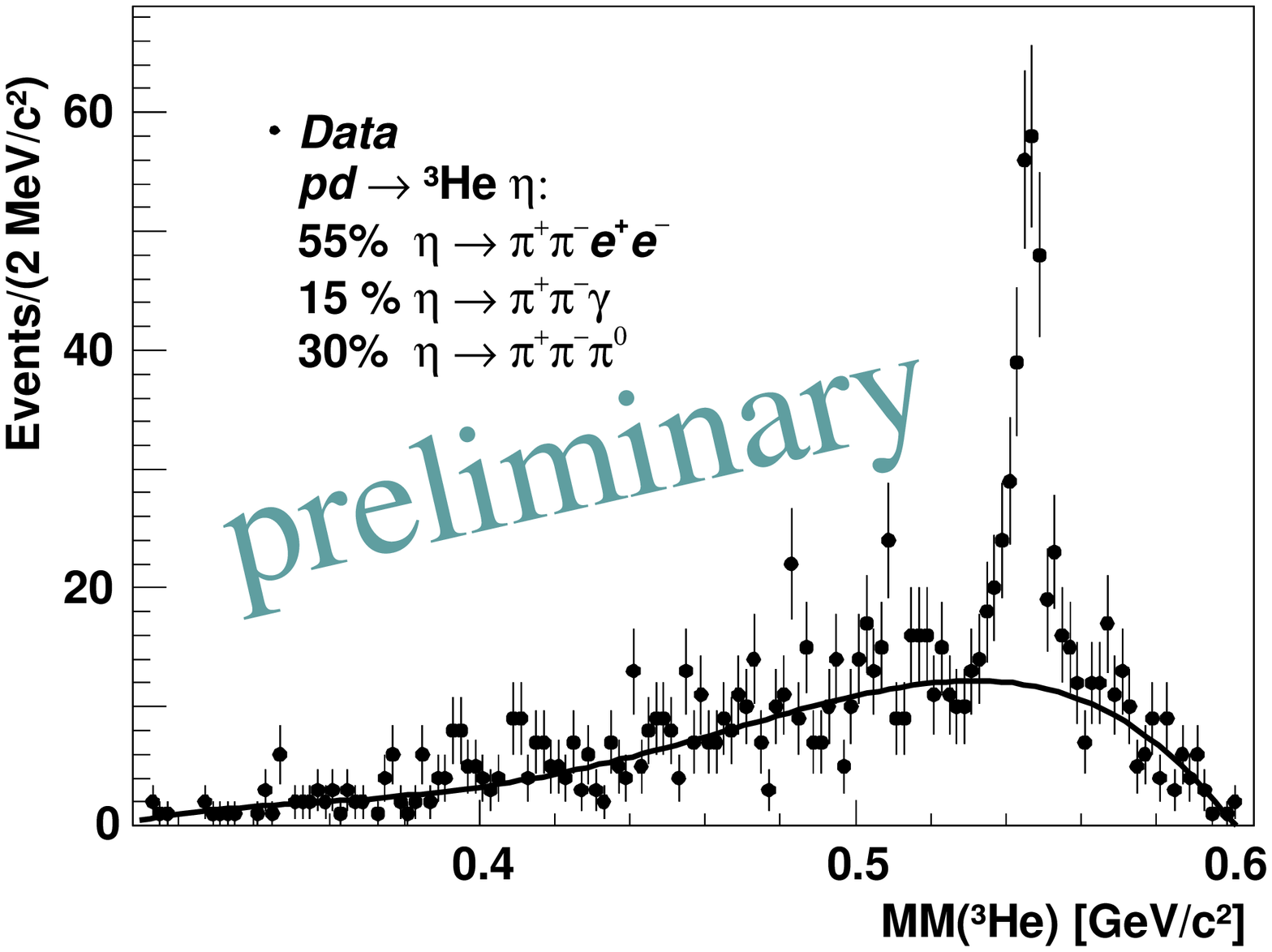}
  \figcaption{Missing mass of ${^3}$He for events compatible with
    $\eta\to \pi^+ \pi^- e^+ e^-$. The line is a polynomial fit to the
    background.}
  \label{fig:pipiee}
\end{center}

The major background contributions come from the $\eta \to
\pi^+\pi^-\gamma$ decay, with $\gamma$ conversion in the beam tube.
Additional background stems from $\eta \to \pi^+\pi^-\pi^0$, due to
internal (Dalitz decay of the $\pi^0$) and external conversion of one
of the photons from $\pi^0\to\gamma\gamma$ decays.  At present, the
analysis does not include any vertex reconstruction which will help to
suppress the background from external conversion.  The remaining
background will be dominated by $\eta \to \pi^+\pi^-\pi^0_{\rm
  Dalitz}$ and can be further reduced by improving the detection
sensitivity for the additional photon.  The overall reconstruction
efficiency is limited by requesting the suppression of the background
and is 5--10~\%.

The measurement of double Dalitz $\eta$ decays, $\eta\to
e^+e^-e^+e^-$ allows the study of the $\eta$ meson form factor for
two timelike virtual photons. In the 2008 $pd$ data, 15 candidate
events have been identified.

The $\eta$  decay into $\pi^0 e^+e^-$  is forbidden from proceeding
via an intermediate  $\pi^0\gamma^\star$ state  due  to
C-conservation. The present  upper limit  for the  branching ratio is
$\le$ 4\E{-5}. The upper limit  would correspond to a  maximum of 400
events  in the 2008 $pd$ run  period.  No  signal is observed  in the
data. Although the analysis is in an early stage, our sensitivity is
already close to the present upper  limit.  A refined  analysis
including a  full kinematic fit will  further suppress the
contributions  from background channels and enhance the sensitivity.

The dominant  mechanism for ${\cal  P}\rightarrow \ell^+\ell^-$
decays within the Standard  Model is a process involving  two virtual
photons and this is additionally suppressed by  helicity
conservation. Recently the interest  in  the  decays  was  revived
due to an observed excess  for $\pi^0\to e^+e^-$  decay branching
ratio \cite{Dorokhov:2007bd}.  For the  $\eta\to e^+e^-$  decay the
best experimental  limit (2.7\E{-5}) comes from CELSIUS/WASA
\cite{Berlowski:2008zz}.

Due to WASA's calorimeter and MDC particle flux limitations, the
maximal rate of $\eta$ mesons from the \pdHeh\ reaction is about 10
$\eta$ events/s.  Therefore the reaction can only be used for studies
of not-so-rare $\eta$ decays.  Further progress towards the
measurement of rare $\eta$ decays can only be made by focusing on the
$pp\to pp\eta$ production reaction.  The reaction has 10--20 times
larger cross section (10 $\mu$b at 1.4 GeV) and the inclusive $pp$
cross section is two times lower than for $pd$ interactions.
Therefore the $\eta$ meson yield will be higher for a given
luminosity and, in addition, one can use up to two times higher
luminosity. A first production beam time of 8 weeks is scheduled for
spring 2010. The goal is to observe about $10^{10}$ $\eta$ decays per
year from then on.

\section{Kaon-pair production}

Over the last decade, exhaustive measurements on kaon-pair production
have been carried out at several COSY facilities. Values of total and
differential cross sections are now available for various isospin
configurations of the entrance and exit channels, \textit{cf.}\
Fig.~\ref{fig:kaonpair}.

\begin{center}
  \includegraphics[width=\linewidth]{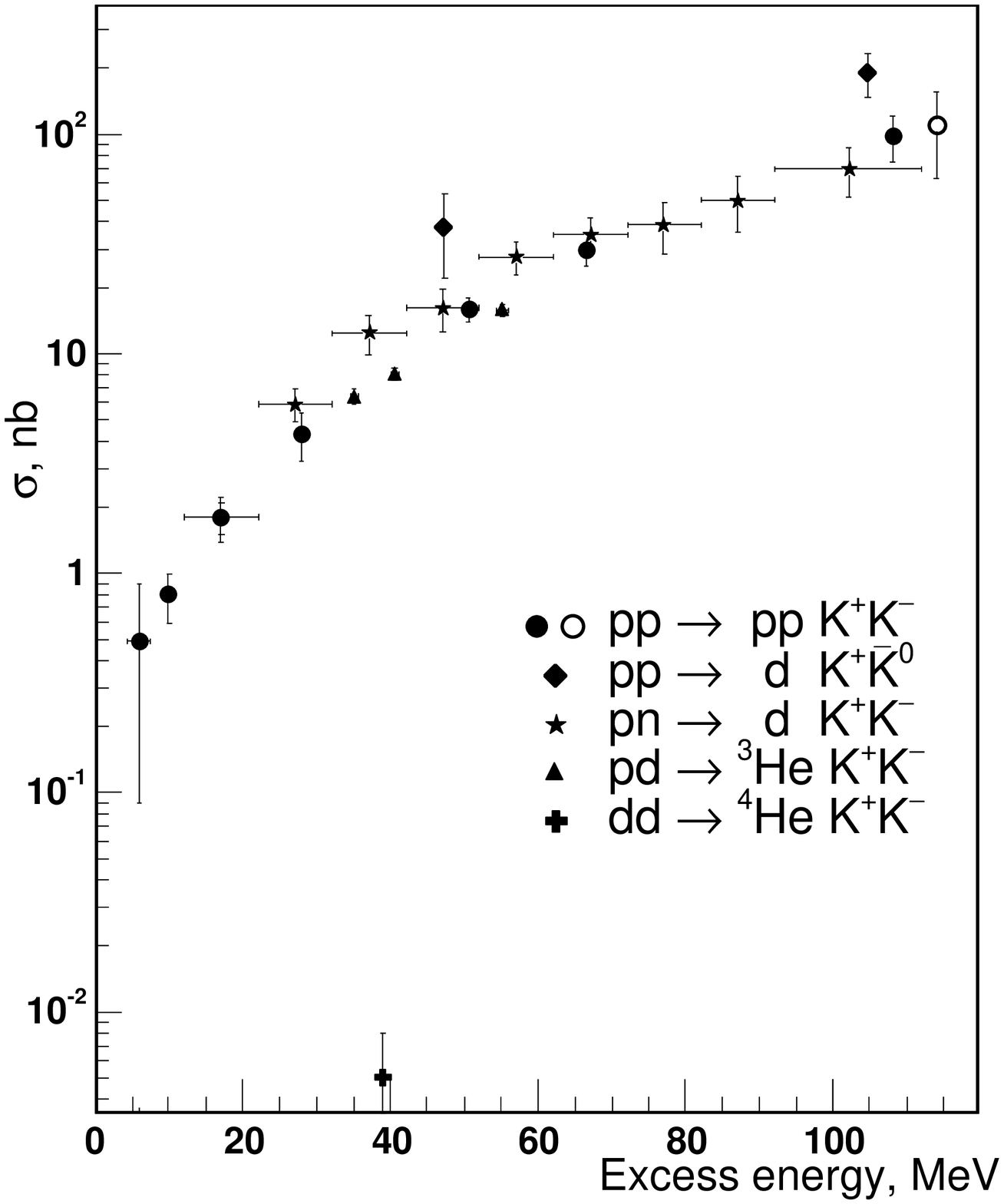}
  \figcaption{\label{fig:kaonpair} World data set on total cross
    sections for kaon-pair production. Closed symbols denote results
    from COSY for the $pp\to ppK^+K^-$, $pp\to dK^+\bar K^0$, $pn\to
    dK^+K^-$, $pd\to{}^3\mathrm{He}\,K^+K^-$, and $dd\to
    {}^4\mathrm{He}\,K^+K^-$ reactions.}
\end{center}

The $pp\to ppK^+K^-$ reaction has been studied by the
COSY-11\cite{Quentmeier:2001ec,Winter:2006vd} and
ANKE\cite{Hartmann:2006zc,Maeda:2007cy} collaborations for excess
energies $\varepsilon$ ranging from 3 to 108\,MeV. Also the
isospin-related $dK^+\bar K^0$ exit channel has been measured at ANKE
for $\varepsilon= 47$ and 105\,MeV\cite{Kleber:2003kx,Dzyuba:2006bj}.
The $pn\to dK^+K^-$ reaction has been investigated at
ANKE\cite{Maeda:2006wv,Maeda:2008mx} using a deuterium cluster-jet as
an effective neutron target. In such measurements, the momentum of
the non-observed proton spectator, and thus the excess energy
$\varepsilon_{K\bar K}=17$--102\,MeV, is reconstructed from the
four-momenta of the remaining detected particles.

These $pN\to (2N)\, K\bar K$ data have been analyzed in terms of
the relevant final-state interactions (FSIs). It turned out that ---
besides the well known interaction of the two outgoing nucleons ---
there is a delicate interplay between the $K\bar K$ and $\bar KN$
FSIs, see {\em e.g.\/} Ref.\cite{Dzyuba:2008pg}. The $K^+N$
interaction is generally believed to be small and has been neglected
in all analyses.

The produced $\bar KN$ systems may provide valuable information about
the much debated antikaon-nucleon interaction strength and, possibly,
about intermediate hyperon states such as the $\Sigma(1385)$ or the
$\Lambda(1405)$. In fact, for the $pp\to dK^+\bar K^0$ reaction the
$\bar Kd$ pairs are found preferentially at small invariant masses;
the partial wave decomposition reveals a $\bar Kd$ $S$-wave
enhancement as compared to $K^+d$\cite{Dzyuba:2008pg}. The same
low-mass enhancement has also been found in the $pn\to dK^+K^-$
data\cite{Maeda:2008mx}. This has been interpreted as evidence for a
strong attractive $\bar Kd$ FSI and the $\bar K^0d$ data are best fit
with a scattering length of $|a_{\bar Kd}| =
1.5$~fm\cite{Dzyuba:2008pg}.  A similar effect has been observed for
the $pp\to ppK^+ K^-$ data. Here the enhancement at low $K^-p$ and
$K^-pp$ masses is best fit with a $\bar{K}p$ scattering length of
$a_{\bar Kp}=(0+1.5i)$~fm\cite{Maeda:2007cy}.

The $K\bar K$ interaction may, in principle, be driven by the light
scalar mesons, $a_0/f_0(980)$, as well as the $\phi(1020)$. In fact,
all data sets with $(K\bar K)^0$ pairs reveal a clear $\phi(1020)$
signal when its production is energetically allowed. The
$a_0/f_0(980)$ have widths of about 50--100 MeV, {\em i.e.\/} much
larger than that of the $\phi(1020)$. They are thus harder to be
found in the $K\bar K$ mass distributions, in particular at COSY
energies where the excess energy is limited to roughly 100~MeV.
Consequently, there is only indirect evidence for scalar-meson
production from the partial-wave decompositions of the $pp\to
ppK^+K^-$ and $pp\to dK^+\bar K^0$ data. These reveal an $S$-wave
dominance in the $K\bar K$
systems\cite{Kleber:2003kx,Dzyuba:2006bj,Hartmann:2006zc}, which may
be interpreted in terms of kaon production {\em via} the $a_0^0/f_0$
and $a_0^+$ channels, respectively.

On top of the above discussed FSIs there is an enhancement at low
$K\bar K$ masses, {\em i.e.\/} between the $K^+K^-$ and
$K^0\bar{K}^0$ thresholds, in the $pp\to ppK^+ K^-$ and $pn\to dK^+
K^-$ data.  This finding has been attributed to
$K^+K^-\rightleftharpoons K^0\bar{K}^0$ charge-exchange
scattering\cite{Dzyuba:2008fi}. This analysis also suggests that
isospin-zero $K\bar K$ production is dominant. Independent of
assumptions about the the details of the intermediate states, a
re-analysis\cite{Silarski:2009yt} of the COSY-11 $pp\to ppK^+ K^-$
data yields an estimate of the $K\bar K$ scattering length:
$|Re(a_{K^+K^-})| = 0.5^{+ 4.0}_{-0.5}$~fm and $Im(a_K^+K^-) = 3.0
\pm 3.0$~fm.

MOMO measured the $pd\to{}^3{\mathrm He}\,K^+K^-$ reaction at three
excess energies, $\varepsilon = 35$, 41 and
55\,MeV\cite{Bellemann:2006xy}. Since the signs of the charges of the
kaons were not identified, these data cannot be used to reliably
isolate the $\bar K^3$He FSI\cite{Grishina:2006cm}.  Also, only upper
limits for the $a_0(980)$ and $f_0(980)$ contributions of 25\% and
10\% could be determined\cite{Grishina:2006cm} while there is a clear
peak in the $K^+K^-$ invariant mass distributions that can be
attributed to a strong $\phi$(1020) contribution..

The total $dd\to{}^4{\mathrm He}\,K^+K^-$ cross section has been
measured at ANKE\cite{Yuan:2009vh}. This reaction is of particular
interest since a $dd\to{}^4{\mathrm He}\,X$ reaction may serve as a
filter for isospin-zero states $X$. However, as seen from
Fig.~\ref{fig:kaonpair}, the total cross section has a value of only
5\,pb, which is several orders below that of all other reactions.
With such a low value, this is the rarest reaction that has so far
been measured at COSY. It should be noted that the total $dd$ cross
section at these energies is about \textbf{eleven} orders of
magnitude larger.

To summarize, there is ample evidence that the antikaon is strongly
attracted to the recoil nucleons in $pN\to (2N)K^+K^-$ reactions.
Only residual effects of the $K^+K^-$ interaction are seen, including
a cusp at the $K^0\bar{K}^0$ threshold. In order to better understand
the dominating $\bar KN$ FSI and, in particular, the role of possible
intermediate $\Sigma(1385)$ and $\Lambda(1405)$ formation a combined
analysis of their strange and
non-strange\cite{Zychor:2005sj,Zychor:2007gf} decays has been
suggested\cite{Wilkin:2008yp}.

In order better to restrict the contribution of the $a_0/f_0$
resonances to the $pd\to{}^3{\mathrm He}\,K^+K^-$ reaction, the WASA
collaboration has measured the $pd\to{}^3{\mathrm
He}\,\eta\pi^0/\pi^0\pi^0$ processes. The data, which are currently
being analyzed, will yield information about the $pd\to{}^3{\mathrm
He}\,a_0/f_0$ cross sections from the detection of the non-strange
strong $a_0\to\eta\pi^0$ and $f_0\to\pi^0\pi^0$ decays. One may hope
that either these mesonic excitations dominate or that they are
negligibly small. In either case the analysis of the $K\bar K$ data
would then be simplified since only one type of production mechanism
({\em viz.\/} scalar mesons or hyperons) would need to be
considered.

\acknowledgments{The author is grateful to P.~Goslawski,
  A.~Kup\'{s}\'{c} and C.~Wilkin for stimulating discussions during
  the preparation of the manuscript.}

\end{multicols}

\vspace{-2mm}
\centerline{\rule{80mm}{0.1pt}}
\vspace{2mm}

\begin{multicols}{2}

\end{multicols}

\clearpage


\begin{thebibliography}{90}

\vspace{3mm}

\bibitem{maier1997} R.~Maier \emph{et al.}, Nucl.\ Instrum.\ Methods Phys.\
  Res.\ Sect.\ A \textbf{390} (1997) 1.

\bibitem{barsov2001} S.~Barsov \emph{et al.}, Nucl.\ Instrum.\
  Methods Phys.\ Res.\ Sect.\ A \textbf{462} (2001) 354.

\bibitem{khoukaz1999} 
  A.~Khoukaz {\it et al.}, Eur.\ Phys.\ J.\ D {\bf 5} (1999) 275.

\bibitem{Buescher:2002zc}
  M.~Buescher {\it et al.},
  Nucl.\ Instrum.\ Meth.\  A {\bf 481} (2002) 378.

\bibitem{Hartmann:2007ks}
  M.~Hartmann {\it et al.}  [ANKE Collaboration],
  Int.\ J.\ Mod.\ Phys.\  A {\bf 22} (2007) 317.

\bibitem{anke-fwd}
  S.~Dymov \textit{et al.},
  Part.\ Nucl.\ Lett.\ \textbf{1} (2004) 40.

\bibitem{ElSamad:2009yu}
  S.~Abd El-Samad {\it et al.}  [COSY-TOF Collaboration],
  Eur.\ Phys.\ J.\  A {\bf 42} (2009) 159
  [arXiv:0906.3095 [nucl-ex]].

\bibitem{AbdelSamad:2006qu}
  S.~Abdel-Samad {\it et al.}  [COSY-TOF Collaboration],
  Phys.\ Lett.\  B {\bf 632} (2006) 27.

\bibitem{Bargholtz:2008ze}
  C.~Bargholtz {\it et al.}  [CELSIUS/WASA Collaboration],
  Nucl.\ Instrum.\ Meth.\  A {\bf 594} (2008) 339
  [arXiv:0803.2657 [nucl-ex]].

\bibitem{Adam:2004ch}
  H.~H.~Adam {\it et al.}  [WASA-at-COSY Collaboration],
  arXiv:nucl-ex/0411038.

\bibitem{mersmann2007} 
  T.~Mersmann {\em et al.}, Phys.\ Rev.\ Lett.\ {\bf 98} (2007) 242301.

\bibitem{Goslawski:2009vf}
  P.~Goslawski {\it et al.},
  arXiv:0908.3103 [physics.acc-ph].

\bibitem{pdg2008} 
  C.~Amsler {\em et al.} [Particle Data Group], Phys.\ Lett.\ B {\bf 667} (2008) 1.

\bibitem{EDDA} 
  M.~Altmeier {\em et al.}, Eur.\ Phys.\ J.\ A {\bf 23} (2005) 351, and
  references therein.

\bibitem{Bashkanov:2007iy}
  M.~Bashkanov {\it et al.},
  Phys.\ Rev.\  C {\bf 76} (2007) 048201
  [arXiv:0708.2014 [nucl-ex]].

\bibitem{Bargholtz:2006gz}
  C.~Bargholtz {\it et al.}  [CELSIUS-WASA Collaboration],
  Phys.\ Lett.\  B {\bf 644} (2007) 299
  [arXiv:hep-ex/0609007].

\bibitem{Berlowski:2008zz}
  M.~Berlowski {\it et al.},
  Phys.\ Rev.\  D {\bf 77} (2008) 032004.

\bibitem{Adolph:2008vn}
  C.~Adolph {\it et al.}  [WASA-at-COSY Collaboration],
  Phys.\ Lett.\  B {\bf 677} (2009) 24
  [arXiv:0811.2763 [nucl-ex]].

\bibitem{Dorokhov:2007bd}
  A.~E.~Dorokhov and M.~A.~Ivanov,
  Phys.\ Rev.\  D {\bf 75} (2007) 114007
  [arXiv:0704.3498 [hep-ph]].

\bibitem{Quentmeier:2001ec}
  C.~Quentmeier {\it et al.},
  Phys.\ Lett.\  B {\bf 515} (2001) 276
  [arXiv:nucl-ex/0103001].

\bibitem{Winter:2006vd}
  P.~Winter {\it et al.},
  Phys.\ Lett.\  B {\bf 635} (2006) 23
  [arXiv:hep-ex/0602030].

\bibitem{Hartmann:2006zc}
  M.~Hartmann {\it et al.},
  Phys.\ Rev.\ Lett.\  {\bf 96} (2006) 242301
  [Erratum-ibid.\  {\bf 97} (2006) 029901]
  [arXiv:hep-ex/0604010].

\bibitem{Maeda:2007cy}
  Y.~Maeda {\it et al.}  [The ANKE Collaboration],
  Phys.\ Rev.\ C {\bf 77} (2008) 015204 [arXiv:0710.1755 [nucl-ex]].

\bibitem{Kleber:2003kx}
  V.~Kleber {\it et al.},
  Phys.\ Rev.\ Lett.\  {\bf 91} (2003) 172304
  [arXiv:nucl-ex/0304020].

\bibitem{Dzyuba:2006bj}
  A.~Dzyuba {\it et al.},
  Eur.\ Phys.\ J.\  A {\bf 29} (2006) 245
  [arXiv:nucl-ex/0605030].

\bibitem{Maeda:2006wv}
  Y.~Maeda {\it et al.},
  Phys.\ Rev.\ Lett.\  {\bf 97} (2006) 142301
  [arXiv:nucl-ex/0607001].

\bibitem{Maeda:2008mx}
  Y.~Maeda {\it et al.},
  Phys.\ Rev.\  C {\bf 79} (2009) 018201
  [arXiv:0811.4303 [nucl-ex]].

\bibitem{Dzyuba:2008pg}
  A.~Dzyuba, M.~Buscher, C.~Hanhart, V.~Kleber, V.~Koptev, H.~Stroher and C.~Wilkin,
  Eur.\ Phys.\ J.\  A {\bf 38} (2008) 1
  [arXiv:0804.3695 [nucl-ex]].

\bibitem{Dzyuba:2008fi}
  A.~Dzyuba {\it et al.},
  Phys.\ Lett.\  B {\bf 668} (2008) 315
  [arXiv:0807.0524 [nucl-th]].

\bibitem{Silarski:2009yt}
  M.~Silarski {\it et al.},
  Phys.\ Rev.\  C {\bf 80} (2009) 045202
  [arXiv:0909.3974 [hep-ph]].

\bibitem{Bellemann:2006xy}
  F.~Bellemann {\it et al.}  [COSY-MOMO Collaboration],
  Phys.\ Rev.\  C {\bf 75} (2007) 015204
  [arXiv:nucl-ex/0608047].

\bibitem{Grishina:2006cm}
  V.~Y.~Grishina, M.~Buscher and L.~A.~Kondratyuk,
  Phys.\ Rev.\  C {\bf 75}, 015208 (2007)
  [arXiv:nucl-th/0608072].

\bibitem{Yuan:2009vh}
  X.~Yuan {\it et al.},
  Eur.\ Phys.\ J.\  A {\bf 42} (2009) 1
  [arXiv:0905.0979 [nucl-ex]].

\bibitem{Wilkin:2008yp}
  C.~Wilkin,
  Acta Phys.\ Polon.\ Supp.\  {\bf 2} (2009) 89
  [arXiv:0812.0098 [nucl-th]].

\bibitem{Zychor:2007gf}
  I.~Zychor {\it et al.},
  Phys.\ Lett.\  B {\bf 660} (2008) 167
  [arXiv:0705.1039 [nucl-ex]].

\bibitem{Zychor:2005sj}
  I.~Zychor {\it et al.},
  Phys.\ Rev.\ Lett.\  {\bf 96} (2006) 012002
  [arXiv:nucl-ex/0506014].

\end{thebibliography}
\end{document}